\documentclass{PoS}
\usepackage{macros_gm2}
\usepackage{amsmath}
\usepackage{subfigure}
\usepackage{afterpage}
\newcommand{\preprintline}{\newline
\rightline{\parbox{5cm}{\large\tt CERN-PH-TH-2015-009}}
}
\title{Towards the physical point hadronic vacuum polarisation from M\"obius DWF}

\ShortTitle{Towards the physical point hadronic vacuum polarisation from M\"obius DWF}

\author{\speaker{Marina Marinkovic}$^{a,b}$, {Peter Boyle$^{c}$, Luigi Del Debbio$^{c}$},
Andreas J\"uttner$^{a}$, Kim Maltman$^{d,e}$, Antonin Portelli$^{a}$\\
~\\
$^{a}$School of Physics and Astronomy, University of Southampton
, Southampton SO17 1BJ, UK\\
$^{b}$CERN, Physics Department, 1211 Geneva 23, Switzerland\\
$^{c}$School of Physics \& Astronomy, University of Edinburgh, EH9 3JZ, UK\\
$^{d}$Department of Physics and Astronomy, York University, Toronto, Ontario, Canada M3J 1P3\\
$^{e}$CSSM, University of Adelaide, Adelaide SA 5005 Australia\\
E-mail: \email{marina.marinkovic@cern.ch,paboyle@ed.ac.uk,luigi.del.debbio@ed.ac.uk, a.portelli@soton.ac.uk,kmaltman@yorku.ca,a.juettner@soton.ac.uk}}

\abstract{We present steps towards the computation of the leading-order hadronic contribution to the muon anomalous magnetic moment on RBC/UKQCD physical point DWF ensembles. We discuss several methods for controlling and reducing uncertainties associated to the determination of the HVP form factor.
~\\
~\\
~\\
~\\
~\\
~\\
~\\
\preprintline
}

\FullConference{
The 32nd International Symposium on Lattice Field Theory\\
23-28 June, 2014\\
Columbia University New York, NY}
\begin{document}

\section{Introduction}
The anomalous magnetic moment of the muon is one of the most precisely measured quantities in particle physics and 
it provides a stringent test of the standard model.
The current experimental measurement and theoretical prediction of the 
muon anomalous magnetic moment $a_{\mu}=(g-2)_{\mu}/2$ show a tension ranging between 3 and 4 standard deviations~\cite{Jegerlehner:2009ry,Hagiwara:2011af}.
With the planned improvement of the experimental precision at Fermilab and at J-PARC,  
further reduction of the theoretical uncertainty is required, in order to be able to resolve the current discrepancy and potentially gain insight into new physics.
The dominant 
uncertainties in the theoretical prediction of $a_{\mu}$  are of hadronic origin: the error of the leading hadronic contribution ($\amhlo$) and the error 
of the hadronic light by light contribution ($\amhlbl$). 
Since it was worked out that the hadronic vacuum polarization (HVP) can be evaluated 
in Euclidean space-time~\cite{Blum:2002ii}, the extraction of $\amhlo$ from first principles has become one of the long-standing goals of lattice community 
(for a recent review see~\cite{{Benayoun:2014tra}}).

The leading hadronic contribution to $a_{\mu}$ is obtained as
\begin{align}
\amhlo=(\frac{\alpha}{\pi})^2 \int_0^{\infty} dQ^2 f(Q^2) \times \hat{\Pi} (Q^2),
\label{eq:amu}
\end{align}
where $\hat{\Pi} (Q^2) = \Pi(Q^2) - \Pi(0)$
denotes the renormalized vacuum polarization tensor and $f(Q^2)$ is a 
known function diverging as $Q^2\rightarrow0$, so that the integrand is strongly peaked at $Q^2\approx m_{\mu}^2/4$. Although the HVP function is accessible from lattice simulations, 
 the total integral is dominated in the low momentum region where relative errors of ${\Pi} (Q^2)$ are enhanced. 
On top of that, the HVP determinations from the lattice show significant pion mass dependence~\cite{Boyle:2011hu}, as well as significant decrease in the signal to noise ratio near $Q^2=0$ as $m_{ud} \rightarrow m_{ud}^{phys}$. 
All this makes the control the systematics in lattice determination of $\amhlo$ challenging even before performing the continuum extrapolation 
and including the finite volume effects, isospin breaking effects etc. 

A traditional approach to extracting $\amhlo$ from lattice computations involves assuming a functional dependence for $\Pi(Q^2)$ and performing a fit of the lattice data to obtain the infrared subtraction $\Pi(0)$. 
Let us note that a number of recently proposed methods allow for direct computation of ${\Pi} (0)$~\cite{deDivitiis:2012vs, Chakraborty:2014mfa}, where the method
of Ref. \cite{deDivitiis:2012vs} requires additional computational effort that was not affordable with DWF fermions at the physical point and 
the application of \cite{Chakraborty:2014mfa} to the DWF ensembles still requires further investigation.
In this work, we stick to the standard approach which requires the extrapolation to $\Pi(0)$.  
We eliminate the systematics from chiral extrapolation by measuring HVP directly at the physical quark mass and 
apply a series of Pad\'e approximants~\cite{Aubin:2012me} 
 in an attempt to asses the systematics introduced by fitting.
\section{DWF action, algorithms and computational strategy}
\label{sec:simpar}
In this preliminary work we use a single ensemble of $2+1$ flavors of M\"obius domain wall fermions~\cite{Brower:2004xi} with physical pion mass and lattice extent $48^3\times 96 \times 24$ (see the left panel of Figure \ref{fig:Pt_vs_Z2}). The ensemble generation has been done with the CPS QCD\footnote{Columbia Physics Systems, {http://qcdoc.phys.columbia.edu/cps.html}} package as a part of the wide spread effort of the RBC-UKQCD collaboration. We quote here only the total number of measured configurations $N_{conf}=88$ separated by  20 molecular dynamics trajectories, the lattice spacing for this ensemble $a^{-1}=1.7295(38) \GeV $ and the renormalization constant for the vector current $Z_V=0.71076(25)$; for more details on the RBC-UKQCD DWF ensembles see~\cite{Blum:2014tka}. 
The measurements of the HVP tensor are using the BAGEL\footnote{ http://www.ph.ed.ac.uk/~paboyle/bagel/Bagel.html} assembler kernel generator and the UKHadron codes. Its recent adaptation includes the hierarchically deflated conjugate gradient (HDCG) algorithm~\cite{Boyle:2014rwa}. 
The usage of the HDCG algorithm on this particular ensemble gave us a $3.5\times$ speed up 
over EigCG and made this computation possible with the available computer resources due to the vastly reduced memory footprint (64 vs. 1500 vectors).

Our strategy for the extraction of the HVP form factor involved the computation of the local-conserved vector two point function, as previously done in~\cite{Shintani:2010ph, Boyle:2011hu}
\begin{align}
\Pi_{\mu\nu}(\Qh)=a^4 \sum_{x} e^{i Q x} \langle J_{\mu}^{C}(x) J_{\nu}^{L} (0) \rangle,
\label{FT}
\end{align}
where $J_{\mu}^{C}$ denotes conserved M\"obius current, $J_{\nu}^{L} $ denotes the local current and $\Qh$
are lattice momenta\footnote{The components of the lattice momenta $\Qh$ are  defined as $\qhm=\frac{2}{a}sin(\frac{a Q_{\mu}}{2})$.}.
The conserved current for the M\"obius DWF action and the details of its implementation are given in~\cite{Boyle:2014hxa,Blum:2014tka}.
The choice of local-conserved (L-C) correlator leaves one Ward identity intact and it is favourable over the computation of the local-local (L-L) and conserved-conserved (C-C) correlator. Namely, C-L still leaves the Ward identity $\qhm\Pi_{\mu\nu}(\Qh)=0$ intact, 
while L-L on the lattice does not satisfy a WI for any of the indices, 
C-C has an additional contact term that needs to be subtracted and in addition C-C is computationally more expensive than the L-C construction for DWF.

We can then relate the HVP form factor with the HVP tensor 
\begin{align}
\Pi_{\mu\nu}(\Qh) = ( \delta_{\mu\nu} \Qh^2 -{\qhm}\qhn)\Pi(\Qh^2).
\end{align}
We work with the diagonal components of the HVP tensor, $\Pi_{\mu\mu}(\Qh)$, and furthermore only inlcude in the analysis momenta 
which have no component in the $\mu$ direction, $\qhm=0$. This selection simplifies the calculation of the HVP form factor $\Pi(\Qh^2)=\frac{\Pi_{\mu\nu}(\Qh)}{\Qh^2}$ and eliminates 
a fraction of the unwanted cutoff effects~\cite{Shintani:2010ph}. 
\section{Point source vs. stochastic sources}
\label{sec:sources} 

\begin{figure}
\centering
\includegraphics[width=0.45\textwidth,height=0.2\textheight,angle=0]{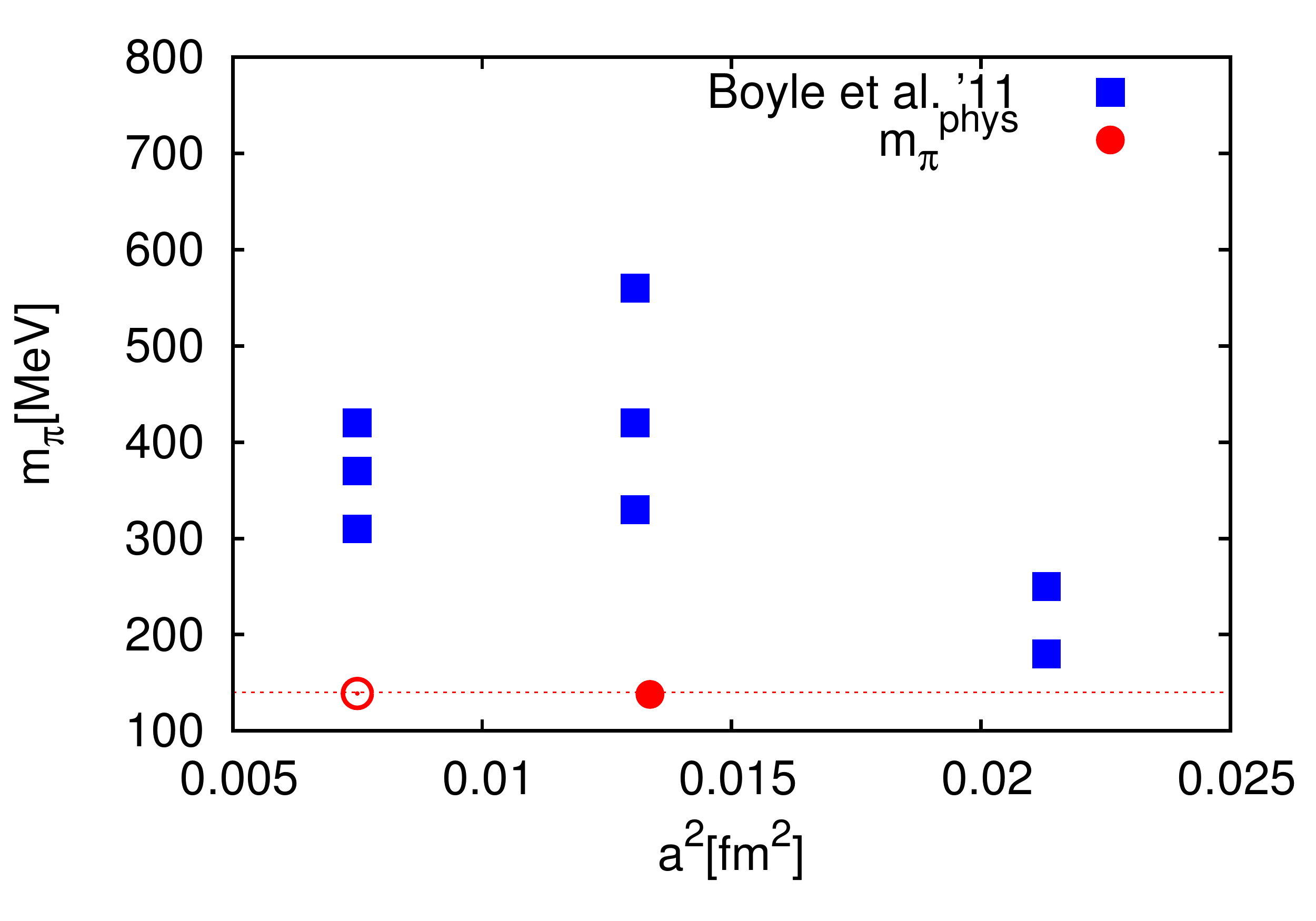}
\includegraphics[width=0.48\textwidth,height=0.2\textheight,angle=0]{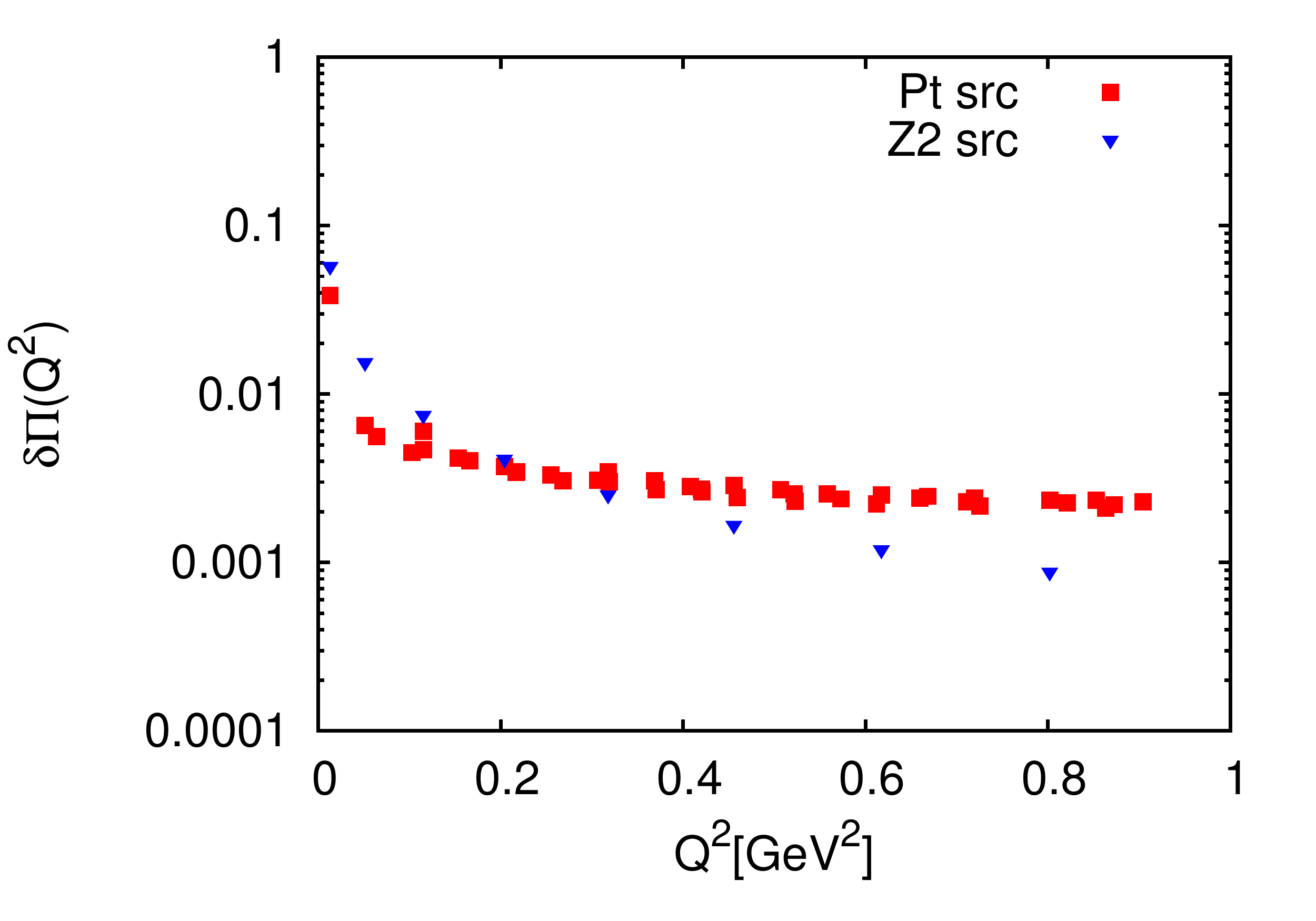}
\caption{ 
Left: overview of the $N_f=2+1$ RBC-UKQCD ensembles used in the HVP programme. Full red circle denotes an
ensemble used in this work, its lattice size being $48^3\times 96 \times 24$ and $m_{\pi}=138\GeV$. Right: comparison of the relative precision of the scalar vacuum 
polarisation computed with a point source (squares) and a $\mathbb{Z}(2)\times\mathbb{Z}(2)$ stochastic source (triangles)
averaged over 12 timeslices on a $48^3\times 96 \times 24$ 
physical point DWF ensemble.
On the y-axis we plot values of $\delta\Pi(\Qh)=\Delta\Pi(\Qh)/\Pi(\Qh)$ obtained by the same number of source positions with both source types, while the 
x-axis reads lattice momenta $\hat{Q}^2$ in $\GeV^2$.}
\label{fig:Pt_vs_Z2}
\end{figure}

Extended stochastic sources have often been proven to be advantageous in the computation of the two point correlators~\cite{Foster:1998vw}, giving the significantly better statistical precision than the point source for same computational cost. 
The 'one-end trick' introduced in~\cite{Foster:1998vw,McNeile:2006nv} and tested for DWF in~\cite{Boyle:2008rh} has been a standard choice of RBC-UKQCD for the calculation of meson two-point and three-point functions. In order to find an optimal setup for the HVP computation on our physical point ensemble of interest, we carry out the same cost comparison of its computation with a $\mathbb{Z}(2)\times\mathbb{Z}(2)$ stochastic wall source~\cite{Boyle:2008rh} with a more conventional computation of the HVP 
with a point source.
We find that after performing the Fourier transform in (\ref{FT}), the volume source performs significantly better than the point source in the large momentum region, but below momenta $\Qh^2\approx 0.2 \GeV^2$ the precision of the point outperforms the volume source we tested. 
This is illustrated in the right panel of 
Figure \ref{fig:Pt_vs_Z2}, where the relative statistical precision of $\Pi(\Qhs)$ for the two choices of the source averaged over 12 timeslices is shown. 
Since this low momentum region is the one where rougly 90\% of the contribution to $\amhlo$ comes from~\cite{Golterman:2014ksa}, we expect the precision of the lattice data in this momentum range 
to be the most important for achieving the desired precision on  $\amhlo$. 
We therefore use point source data for the analysis presented in the following. 
It is possible that the noise characteristics may differ if different analysis methods are used~\cite{Lehner:XX} and that will be the subject of further study.
\section{Pad\'e fits}

\label{sec:pades}
The evaluation of the integral (\ref{eq:amu}) requires asuming a functional dependence of $\Pi(\Qh^2)$. In the first 
attempts to obtain $\amhlo$ from the lattice, many of the functional forms used to date were based on Vector Meson Dominance (VMD).
Recent studies, however, provide strong evidence that this approach introduces uncontrolled systematics~\cite{Aubin:2012me,Golterman:2013vca}. 
We use in this work a series of Pad\'e fits
\begin{align}
\Pi_{[N,D]} (\Qh^2)=\frac{\sum_{n=0}^{N} a_n \Qh^{2n}}{1+\sum_{n=1}^{D} b_n \Qh^{2n}}
\label{pades}
\end{align}
to test the stability of the fit for different number of fit parameters, without relying on VMD. We use the physical pion mass data set described in Section \ref{sec:simpar} and perform [1,1], [2,1] and [2,2] Pad\'e fits which have 3, 4 and 5 free parameters respectively. 
These lie in the sequence of Pad\'es proposed in Ref.~\cite{Aubin:2012me}, known to be such that the full sequence converges to the actual polarization for any compact region in the complex plane excluding the cut along the negative real axis. 

\begin{figure}[t!]
\begin{center}
\subfigure{
\includegraphics[scale=0.25,angle=0]{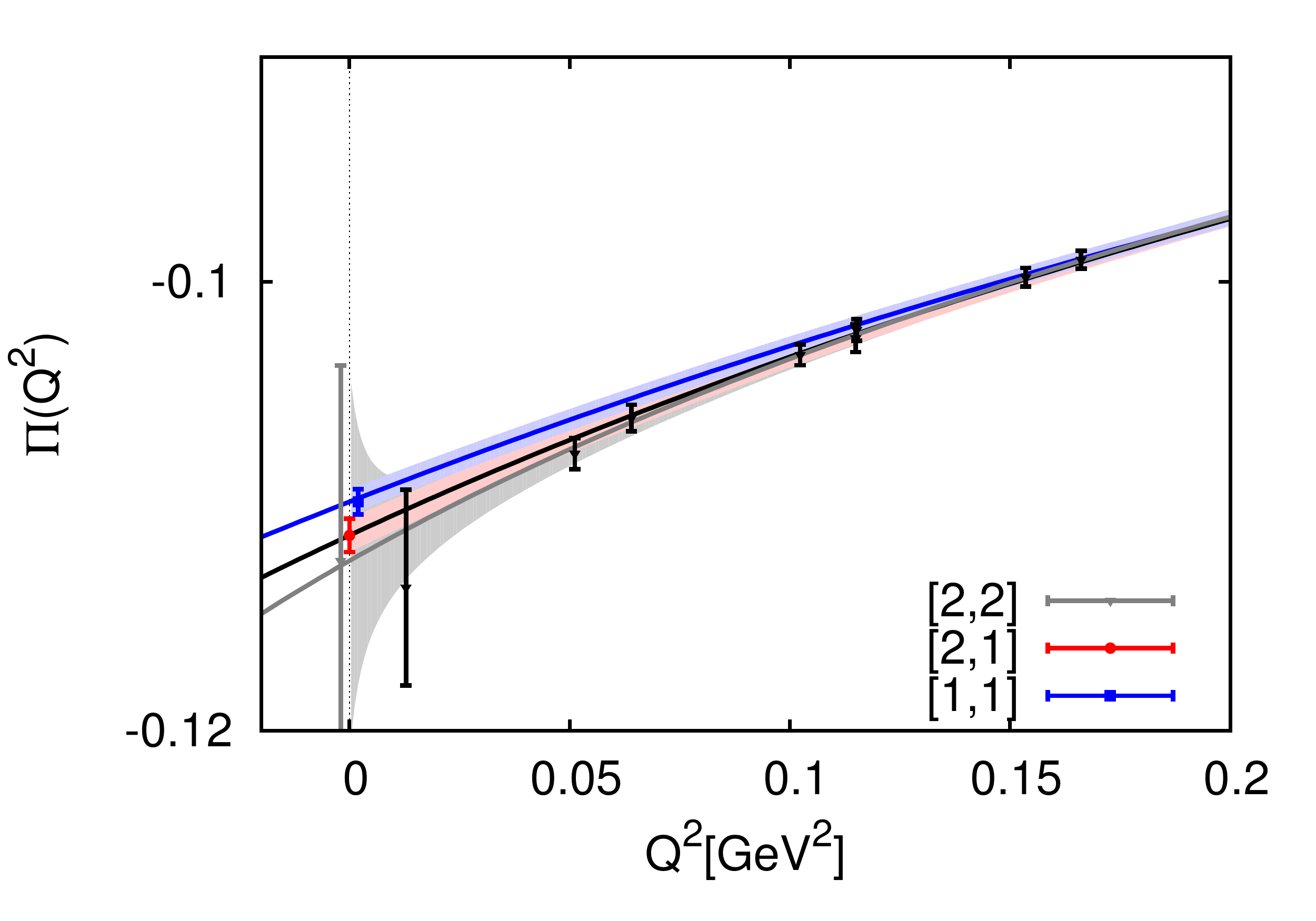}
}
\hspace*{8.0mm}
\subfigure{
\includegraphics[scale=0.25,angle=0]{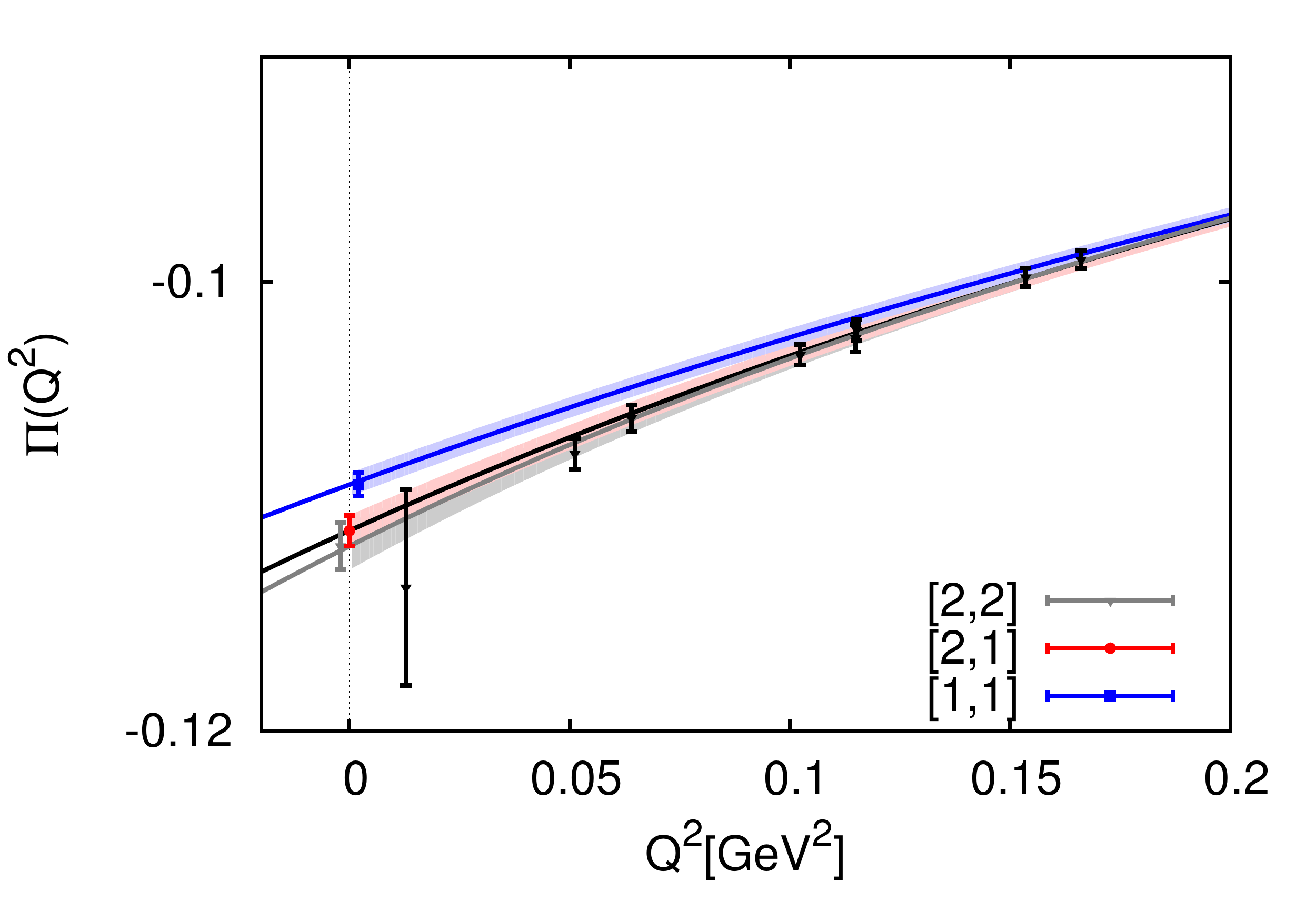}
}
\subfigure{
\vspace*{-8mm}
\includegraphics[scale=0.25,angle=0]{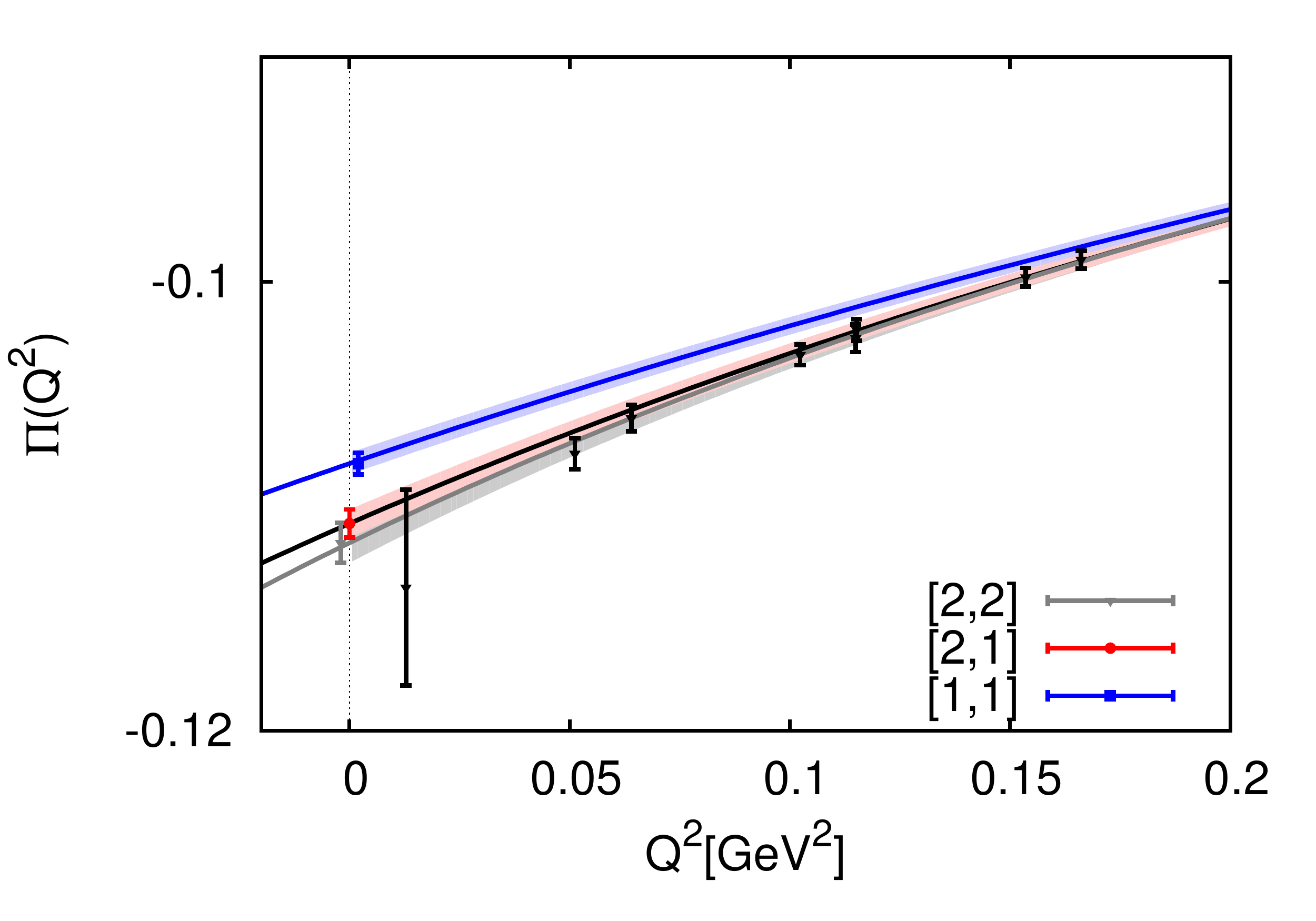}
}
\hspace*{8.0mm}
\subfigure{
\includegraphics[scale=0.25,angle=0]{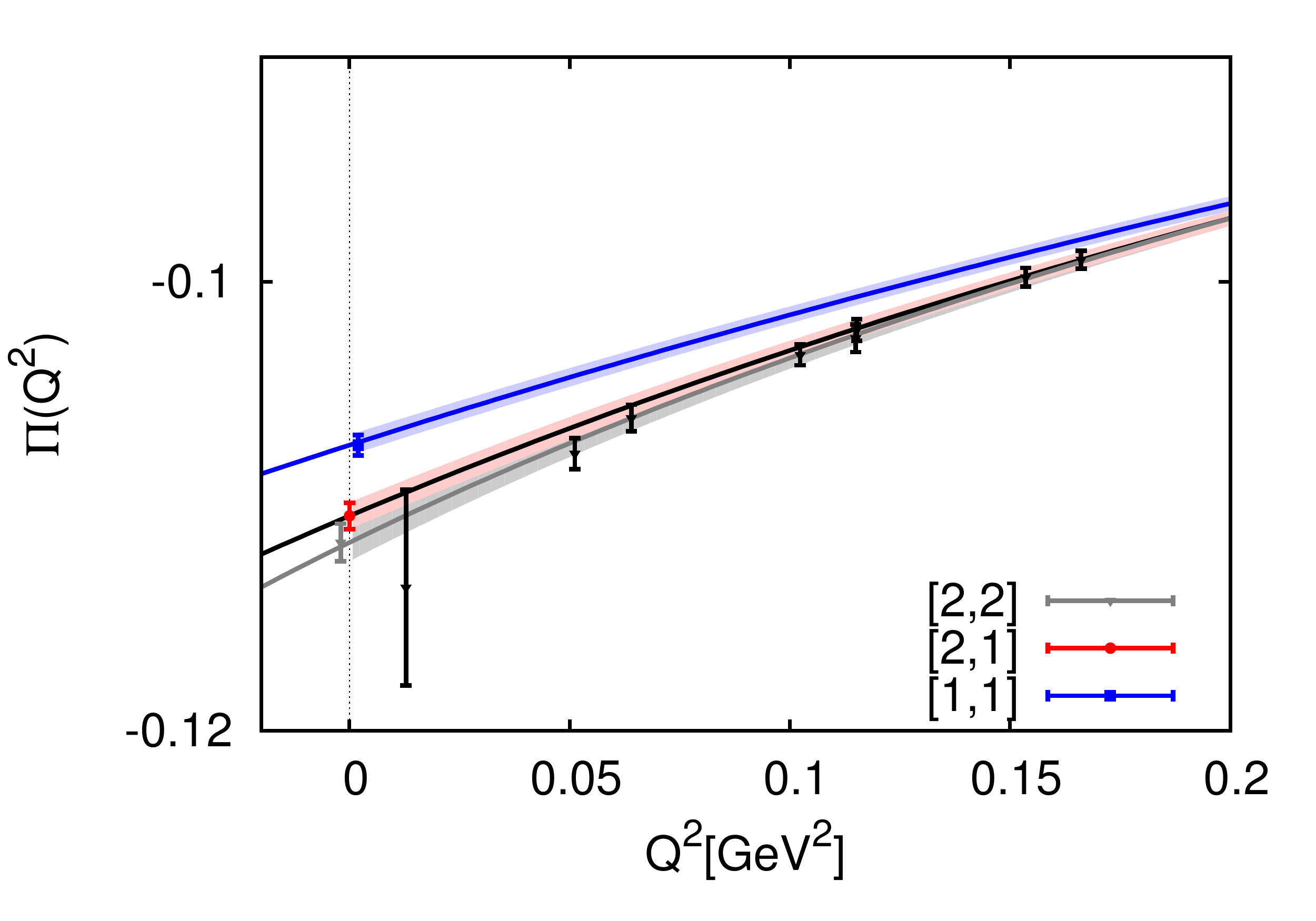}
}
\end{center}
\vspace*{-5mm}
\caption{Using different Pad\'es for extrapolating $\Pi(\Qh^2)$  to $\Qh^2=0$ on a $48^3\times96\times24$ lattice with $a^{-1}=1.7295\GeV$ and $m_{\pi}=0.138\GeV$. Upper left plot includes in the fit 
 momenta up to $\Qh^2<Q_C^2=1.5\GeV^2$, upper right includes $\Qh^2<Q_C^2=2\GeV^2$, lower left plot includes $\Qh^2<Q_C^2=2.5\GeV^2$ in the fits and lower right fits momenta up to $\Qh^2<Q_C^2=3\GeV^2$.}
\label{fig:Pi0}
\end{figure}

Let us first discuss the extrapolation of  $\Pi(\Qh^2)$  to $\Qh^2=0$. In Figure \ref{fig:Pi0} we show different orders of Pad\'es applied to a fitting interval $[0,Q_C^2]$. 
For the lowest cut value we took $Q_C^2=1.5\GeV^2$, 
the fit [2,2] gives large uncertainities and the extrapolated values of $\Pi(0)$ from all attempted fits are compatible with each other. For the remaining three values of $Q_C^2$ both Pad\'es [2,1] and [2,2] give good $\chi^2$, while Pad\'e [1,1] gives $\Pi(0)$ more than one sigma away from the higher order Pad\'es.  We ilustrate this in the remaining three panels of Figure \ref{fig:Pi0}.
For cut values $Q_C^2=2\GeV^2;2.5\GeV^2$ {and} $3 \GeV^2$ the extrapolated $\Pi(0)$ values from [2,1] and [2,2] Pad\'es are compatible with each other and one might be tempted to consider [2,1] a fit of choice, 
since it has less free parameters then [2,2] and therefore gives better statistical precision.
We will see in the following section that using lower order Pad\'es in combination with higher cuts in the momenta  $Q_C^2$  introduces systematics one would like to avoid.
\section{Including large $\Qh^2$ in the Pad\'e fits}
\label{sec:syst}
\begin{figure}[t!]
\begin{center}
\subfigure{
\includegraphics[scale=0.25,angle=0]{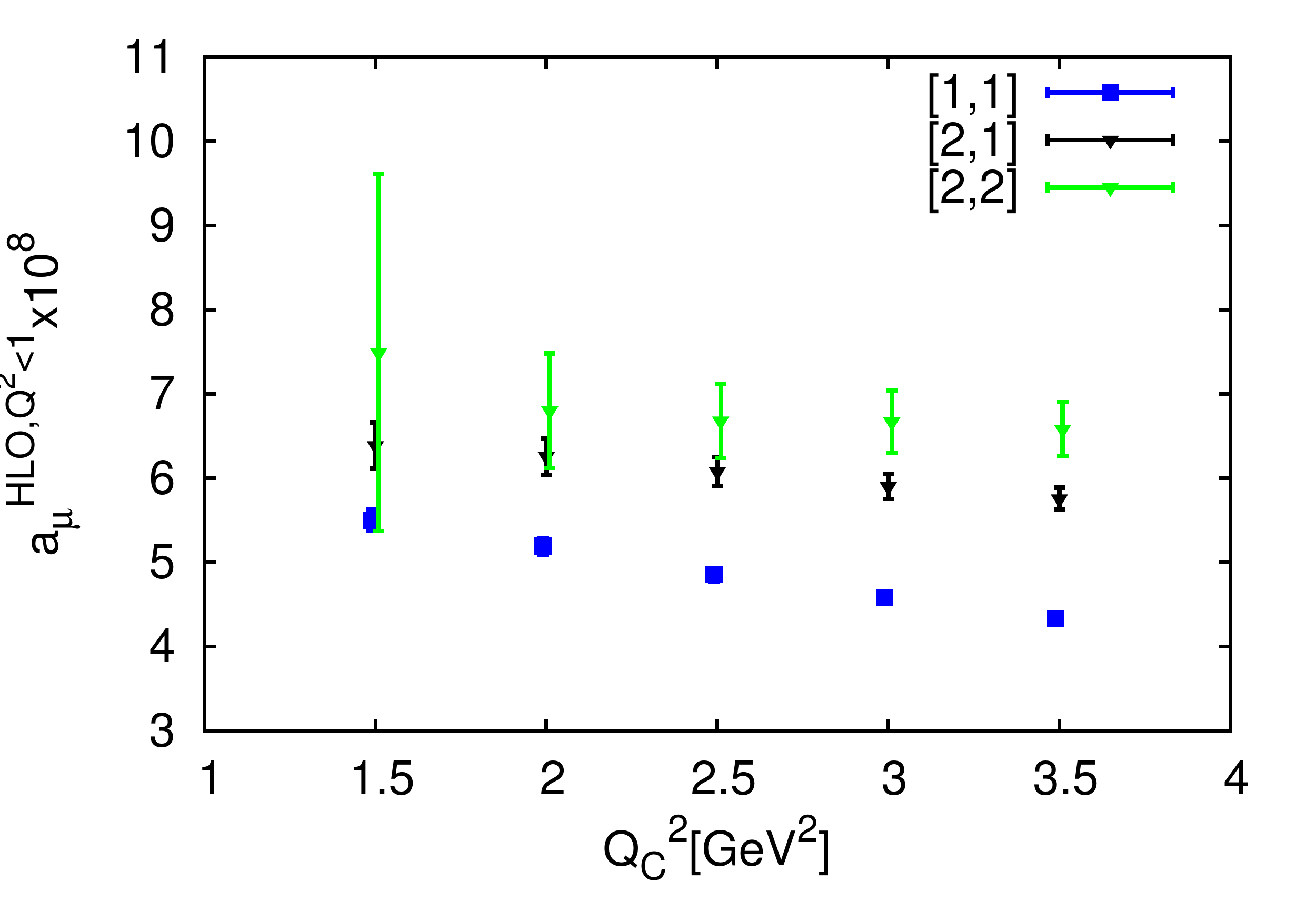}
\label{sim_Qc}
}
\hspace*{8.0mm}
\subfigure{
\includegraphics[scale=0.25,angle=0]{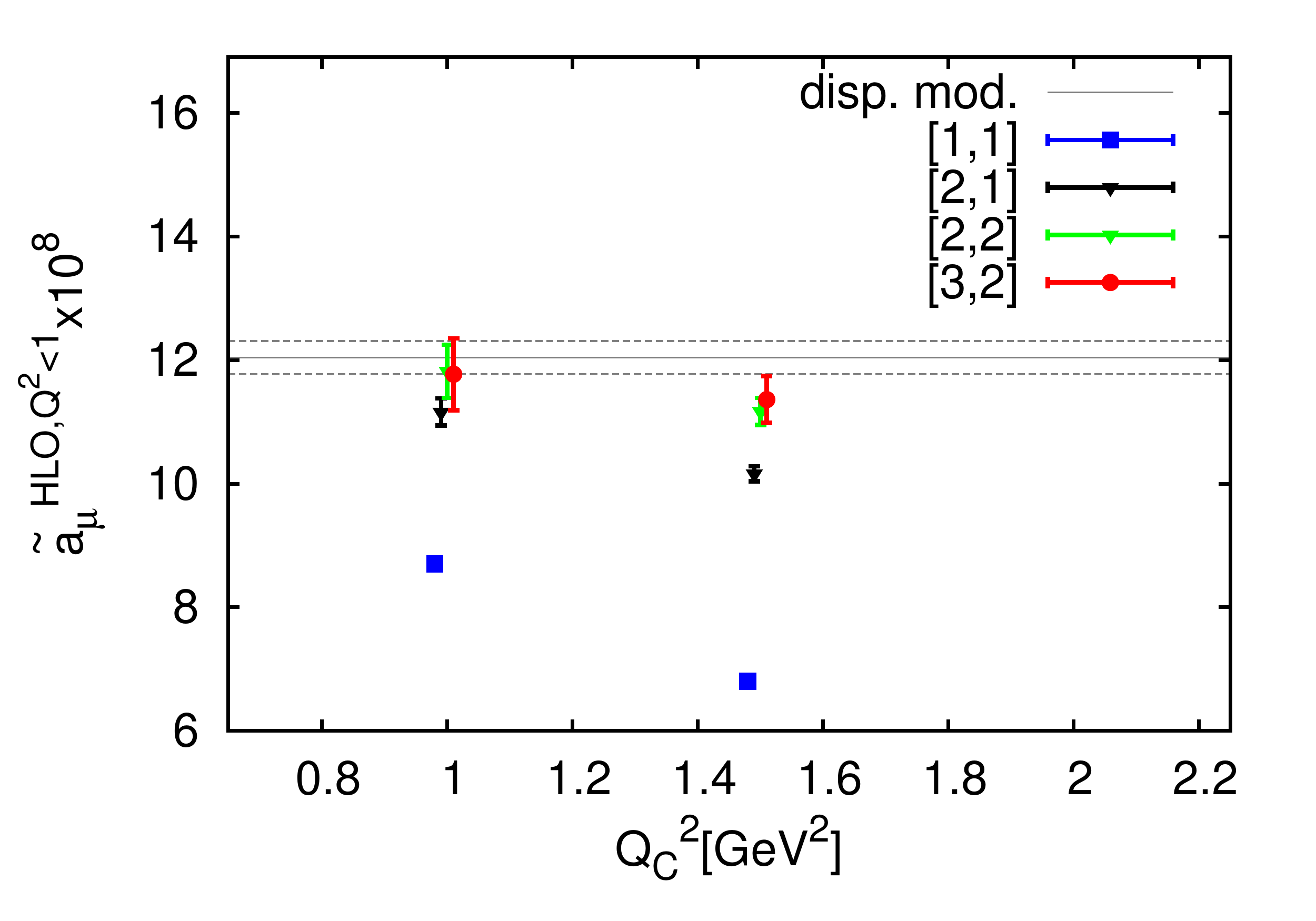}
\label{disp_Qc}
}
\end{center}
\vspace*{-5mm}
\caption{Left: contribution to $\amhlo$ when the corresponding Pad\'e fits defined in 
eq. (4.1) are performed in the momenta region 
$\Qh^2 \in [0,Q_C^2]$ for different values of maximal $Q_C^2$ included in the fit. Right: Result for the $I=1$ non-strange polarization function obtained from the dispersive model defined in~\cite{Golterman:2013vca}. The plotted contributions to $\tilde{a}_{\mu}$ from the momenta region $ [0,1]\GeV^2$ for different Pade approximants$^{\tiny{3}}$
$[N,D]$ are taken from Table 1 in Ref.~\cite{Golterman:2013vca}.}
\label{fig:Qc}
\end{figure}

Including momenta up to higher values of $Q_C^2$ in Pad\'e fits with small number of parameters is yet another potential source of systematic uncertainty. 
This was first observed for the dispersive model constructed for studying the uncertainities associated with different fits to lattice data for the vacuum polarization~\cite{Golterman:2013vca}. 
Namely, when the maximal momenta included in the fits of the $I=1$ polarization data has been changed from $Q_C^2=1\GeV^2$ to $Q_C^2=1.5\GeV^2$, the value of the relevant vacuum polarization contribution is undershooting its true value, which in this particular model can be directly extracted from the hadronic $\tau$-decays data (see right panel of Figure \ref{fig:Qc}).
We probe on our data set several values of $Q_C^2$ together with the Pad\'e fits used in the previous section.
In the left panel of Figure \ref{fig:Qc} we show the contributions $\amhlo(\Qh^2<1\GeV^2)$ resulting from different Pad\'e fits
on our data set.
Only higher order Pad\'es seem to give consistent contributions to $\amhlo$ for higher values of $Q_C^2$, as already suggested by the dispersive model study. 
In particular, 
on our ensemble and with the current statistics, both [1,1] and [2,1] Pad\'e are not acceptable for the values of $Q_C^2>1.5\GeV^2$. Even though the values of $\amhlo(\Qh^2<1\GeV^2)$ obtained by [2,1] Pad\'e function look compatible for different values of $Q_C^2$, note that the plotted values are correlated and once the correlations are subtracted, the spread comes out to be larger than the corresponding estimates of the statistical errors which favours either using higher order Pad\'es such as [2,2] or some kind of hybrid methods along the lines of~\cite{Golterman:2014ksa} where  the fitting interval is significantly shorter.
\footnotetext{Note that Pade approximants $[N,D]$ plotted in the right panel of Figure 3  correspond to [N-1,D] in the notation of Ref.~\cite{Golterman:2013vca}.}
\section{Summary}
We are measuring the leading order hadronic contribution to $a_{\mu}$ with $2+1$ flavors of DWF fermions at the physical point. 
At this point of the calculation and on the studied physical point data set, we find that the point source gives better relative precision than the stochastic source 
with the one-end trick  
for the momenta $\Qh^2<0.2\GeV^2$, which is the region 
that gives most of the total $\amhlo$ contribution.
We perform fits using Pad\'e approximants, which constitute a model-independent parametrisation of the HVP form factor, and test how changing the number of free parameters and the fitting range affects the 
infrared subtraction and the evaluation of the integral (\ref{eq:amu}). 
The Pad\'es we employ lie in the sequence proposed in Ref.~\cite{Aubin:2012me} and are guaranteed to converge to the actual polarization, but the systematics associated with the use of particular order Pad\'es with particular choices of $Q_C^2$ remain to be explored in more detail. Progress stabilizing the fit results may also be possible using the conformal variable polynomial fit forms discussed in Ref.~\cite{Golterman:2013vca}.
The contributions to $\amhlo$ we are getting in Section \ref{sec:syst} are in the same ballpark as previous HVP computations, but with large statistical 
uncertainities. Therefore, already with this computation on a single ensemble we can see that more sophisticated methods are necessary to extract the value 
with the precision comparable to the experimental one. 
\section{Acknowledgements}
This work is part of a brother programme of RBC-UKQCD collaboration and the authors thank its members and in particular 
members of DWF-GM2 working group for useful discussions. 
The research leading to these results has received funding from the
European Research Council under the European Community's Seventh
Framework Programme (FP7/2007-2013) ERC grant agreement No 279757. The authors also acknowledge STFC grants ST/J000396/1 and ST/L000296/1.
The calculations reported here have been done on DIRAC Bluegene/Q computer at the University of Edinburgh's Advanced Computing Facility.


\end{document}